\documentclass[prl,twocolumn,aps,superscriptaddress,footinbib]{revtex4-2}
\usepackage{xspace,amsmath,amsfonts,amssymb,amsbsy}
\usepackage{ntheorem}
\usepackage{graphicx,epstopdf}
\usepackage{bbm}
\usepackage{footnote}
\usepackage{braket}
\usepackage[usenames,dvipsnames]{xcolor}
\usepackage[breaklinks=true]{hyperref}
\hypersetup{
colorlinks   = true, %Colours links instead of ugly boxes
urlcolor     = blue, %Colour for external hyperlinks
linkcolor    = blue, %Colour of internal links
citecolor    = red %Colour of citations
}
\usepackage{soul}
\usepackage{ulem}
\makeatletter

\begin{document}
%\title{Dispersion engineering  and soliton generation via cavity-mediated interaction in a thermal gas}

\title{ Solitons  in arbitrary dimensions stabilized by photon-mediated interactions}

\author{Haoqing Zhang}
\affiliation{JILA, NIST and Department of Physics, University of Colorado, Boulder, Colorado 80309, USA}
\affiliation{Center for Theory of Quantum Matter, University of Colorado, Boulder, Colorado 80309, USA}
\author{Anjun Chu}
\affiliation{JILA, NIST and Department of Physics, University of Colorado, Boulder, Colorado 80309, USA}
\affiliation{Center for Theory of Quantum Matter, University of Colorado, Boulder, Colorado 80309, USA}
\author{Chengyi Luo}
\affiliation{JILA, NIST and Department of Physics, University of Colorado, Boulder, Colorado 80309, USA}
\author{James K. Thompson}
\affiliation{JILA, NIST and Department of Physics, University of Colorado, Boulder, Colorado 80309, USA}
\author{Ana Maria Rey}
\affiliation{JILA, NIST and Department of Physics, University of Colorado, Boulder, Colorado 80309, USA}
\affiliation{Center for Theory of Quantum Matter, University of Colorado, Boulder, Colorado 80309, USA}

\date{\today}
\begin{abstract}
We propose a scheme to generate solitons
in arbitrary dimensions, in a matter-wave interferometer, without the need of quantum degeneracy. 
%Solitons are .
In our setting,  solitons
%, or waves that maintain their shape while propagating, 
emerge by balancing the single-particle dispersion with engineered cavity-mediated exchange interactions between two wave packets, which, at the appropriate conditions, remain bound to each other and dispersion-free.
% In our setting, solitons can be prepared in an arbitrary coherent superposition, and,any dimensions by the use of appropriate drives,
For detection in thermal gases, we propose an interferometric probing scheme instead of traditional time-of-flight imaging. 
%This system opens an avenue for quantum-enhanced matter-wave interferometry relevant for both long evolution times in outer space and in compact devices.
\end{abstract}
\maketitle

\textit{Introduction.--} 
Solitons or self-enforcing waves that propagate without spreading are a fundamental manifestation of nonlinear physics across diverse systems, from classical water waves to optical pulses in fiber communications. The realization of Bose-Einstein condensation (BEC) in weakly interacting atomic gases opened new avenues for exploring quasi-one dimensional (1D) matter-wave solitons and nonlinear phenomena \cite{Lewenstein1999,denschlag2000generating,khaykovich2002formation,strecker2002formation,Eiermann2004}. 
%They include,  quasi-one dimensional (1D)  stable bright (dark) solitons due to the competition between the atom's single-particle dispersion and their attractive (repulsive) interactions~
In higher dimensions (2D and 3D), stable soliton formation is more complicated, due to the presence of dynamical or thermodynamical instabilities and interaction-induced collapse. Only recently, stable bright 2D solitons have been realized \cite{Bakkali-Hassani2021,Chen2021}.
%realized using a two-component planar Bose gas~\cite{Bakkali-Hassani2021} or via a shape-controlled modulational instability~\cite{Chen2021}. 

Atomic solitons have also been identified as promising candidates for inertial sensing,  due to their dispersionless character over long integration times, and their ability to mitigate spatial inhomogeneities during pulse sequences \cite{polo2013soliton,cuevas2013interactions,helm2015sagnac,grimshaw2022soliton,helm2012bright,marchant2013controlled,nguyen2014collisions,mcdonald2014bright}. However, the need to prepare atoms in a BEC, imposes practical limits for sensing tasks. 

In this work, we discuss a way to generate solitons in thermal atoms, and in arbitrary dimensions via exchange interactions between momentum states in an optical cavity.  We use exchange interactions to open a gap in the many-body spectrum to protect spin alignment and gain robustness against single-particle inhomogeneity~\cite{rey2008many,deutsch2010spin,norcia2018cavity,smale2019observation,deutsch2010spin,luo2023cavity,niu2025many,lewis2021cavity,young2024observing,young2024time}, 
as reported in prior work ~\cite{luo2023cavity}, but in addition, we demonstrate that, for specific values of the exchange interaction energy, cavity-mediated interactions can also prevent the overall dispersion of the atoms and lead to a soliton made of two overlapped wave packets.  

%Exchange interactions have been shown to prevent the spatial separation of two interfering wave packets, keeping them together in a way akin to the M{\"o}ssbauer suppression of recoil of atomic nuclei~\cite{pound1960variation,luo2023cavity}. However, before  the interaction could not  prevent the overall dispersion of the atomic ensemble~\cite{luo2023cavity}

Our implementation has some similarities to theoretical ideas based on spin-orbit coupling (SOC)~\cite{dalibard2011colloquium,lin2011spin,wang2012spin,huang2016experimental} to stabilize solitons with contact interactions in 2D~\cite{sakaguchi2014creation} and 3D~\cite{zhang2015stable} systems, but exhibit several distinctive features: our solitons possess fringes that depend on the coherence between different momentum states; they do not inherently require a BEC and can emerge in a thermal gas; and they can manifest in both 1D, 2D, and 3D systems. Furthermore, the same exchange interactions that generate the solitons can entangle the momentum states, paving the way for the application of this system to quantum-enhanced interferometry~\cite{greve2022entanglement}. 
 
\begin{figure}
\includegraphics[width=0.9\columnwidth]{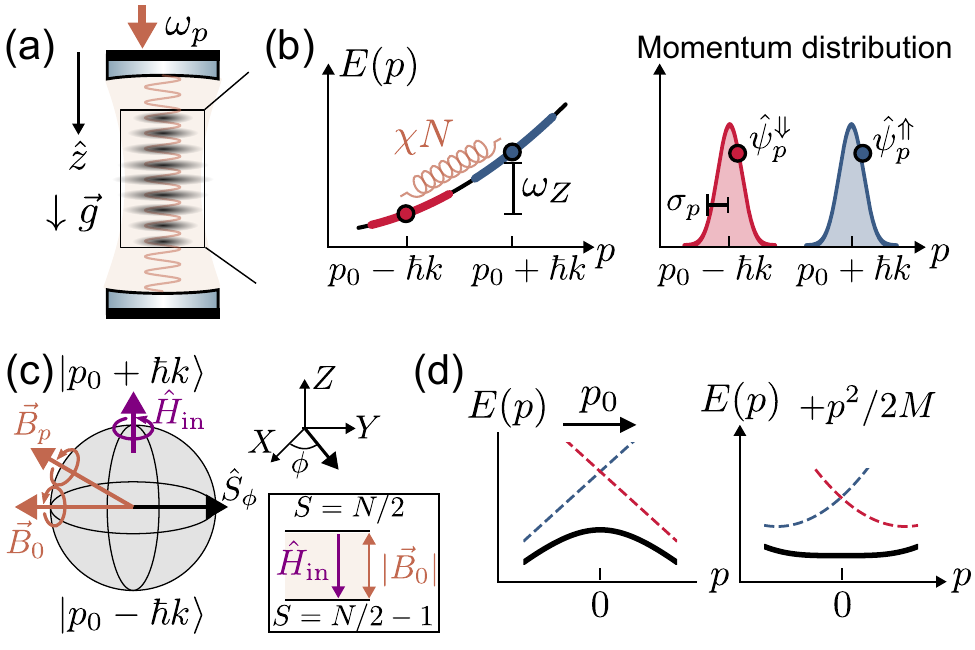}
\caption{\label{fig:scheme}
(a)   Schematic of the interferometry protocol in a cavity.
%After velocity selection, a Bragg pulse prepares $N$ atoms inside an optical cavity in a coherent superposition of 
(b) Cavity-mediated interactions between atoms in the two wave packets, each with average momentum $p_0-\hbar k$  and $p_0+ \hbar k$, momentum spread $\sigma_p$,  and separated by an energy difference $\omega_Z$, create an energy gap, $\chi N$, which keeps the wave packets bound. 
(c) At the mean-field level, for $\chi <0$,   the  Bloch vector of an individual atom (initially pointing on the equator, along an initial angle $\phi$, $\hat{S}_{\phi}$ (black arrow), processes around the effective magnetic field (orange arrows) $\vec{B}_{p_r}= \vec{B}_0 + 2k p_r/M \hat{Z}$, with contributions from the self-generated field and kinetic energy (purple) and with $\vec{B}_0=\chi N \hat{S}_{\phi}$.
(d) At the optimal interaction strength,  in a frame moving with momentum $p_0$,  the modified Doppler dispersion by the exchange interactions plus the free dispersion  (dashed lines) generates a net flat dispersion (black solid line), enabling the formation of a soliton. 
%The indicates the linear contribution of the Doppler  shift (left) which adds to the quadratic  dispersion (right)  for the case of  free particles.}
% (centered at blue and red dots)  
}
\end{figure}

\textit{Model.--} We consider the experimental setup recently implemented in Refs.~\cite{luo2023cavity,luo2025hamiltonian,luo2024realization} and shown in Fig.~\ref{fig:scheme}(a). In this setting, an ensemble of $N_0$ laser-cooled $^{87}\rm{Rb}$ atoms with mass $M$ is placed inside a vertical optical cavity along the $\hat{z}$ direction. 
The initial thermal distribution of the motional degrees of freedom can be described by the Wigner function $f(\mathbf{r},\mathbf{p})\propto  e^{-\beta (\frac{\mathbf{p}^2}{2M}+U(\mathbf{r}))}$ with a characteristic thermal width $\sigma_T=\sqrt{\beta /(M \omega ^2)}$, where  $\beta=1/{k_B T}$ is the inverse temperature. $U(\mathbf{r})$ is the 3D trapping potential energy. 
The trap along $\hat{z}$ is then released, allowing atoms to fall freely under the gravitational acceleration,  $\vec{g}$, with only confinement along radial direction $R$ and prepared in a well-defined momentum state $(p_0-\hbar k) \hat z$ via velocity selection, without changing the atom's internal quantum state. Here  $k=2\pi/\lambda$ and  $\lambda=780$ nm for Rb atoms~\cite{luo2023cavity,luo2025hamiltonian,luo2024realization}.
This process transforms the Wigner function for the $N$ selected atoms into $\tilde{f}(\mathbf{r},\mathbf{p})\propto e^{-(p-(p_0-\hbar k))^2/2\sigma_p^2} \tilde{F}(z,R,p_{R})$, featuring a much narrower momentum spread $\sigma_{p} \ll \hbar k\ll \sigma_T$ along $\hat{z}$. Here $\tilde{F}$ describes the part of the Wigner function that depends on the remaining degrees of freedom.

A two-photon Bragg transition, applied after the momentum selection, coherently couples an atom in momentum $\ket{p -\hbar k}$ to another momentum state $\ket{p+\hbar k}$ within the same internal level, enabling the preparation of arbitrary coherent superpositions ($\hat {\psi}_{p-\hbar k}\to \cos(\theta/2)\hat {\psi}_{p-\hbar k}+ \sin(\theta/2) \hat {\psi}_{p+\hbar k}$) which generate a density grating in position space (Fig.~\ref{fig:scheme}(a)). Here we have defined $\hat {\psi}_{p -\hbar k}$ to be the annihilation momentum operator of a particle with momentum $p-\hbar k$ and $\theta$ is the pulse area. 

Cavity-mediated interactions between momentum states can be generated by driving the cavity with a laser with frequency $\omega_p$, detuned from the atomic transition --- with frequency $\omega_a$ --- by $\Delta_a=\omega_a-\omega_p$, and from the cavity resonance, $\omega_c$, by $\Delta_c=\omega_c-\omega_p$. 
% This drive creates an optical potential within the cavity that is modulated by the falling atomic density grating, which oscillates at a frequency $\omega_z=2\hbar k  \bar{p} /M$ determined by the average kinetic energy difference between the two momentum states $\bar{p}+p \pm \hbar k$~\cite{luo2023cavity,wilson2024entangled} with $\bar{p}=p_0+\hbar k$.
By adiabatically eliminating the atomic excited state and the cavity field, one can obtain an effective Hamiltonian within the ground state manifold that governs only the motional degrees of freedom~\cite{shankar2019squeezed,luo2023cavity,wilson2024entangled}. 

For simplicity, we will assume a single realization of the experiment, in which the selected atoms, before the Bragg pulse, occupy a set of momentum states which we label as $\{p_1,p_2,\dots p_N\}$, where $p_r$ describes a particle with momentum $p_0 -\hbar k + p_{r}$ sample from the Wigner function, $\tilde{f}(\mathbf{r},\mathbf{p})$. The initialization protocol allows us to focus on momentum states within the region $[p_0-2 \hbar k, p_0+2\hbar k]$. In a frame moving with momentum $p_0$, we define the basis states $\ket{\Downarrow_p}_n\equiv \ket{p-\hbar k}_n$ and $\ket{\Uparrow_p}_n\equiv \ket{p+\hbar k}_n$ for $p\in[-\hbar k, \hbar k]$, and introduce the pseudospin-1/2 operator  $\hat{s}_{p_n}^{\eta}=\sum_{\alpha,\beta=\Uparrow,\Downarrow}(\hat{\psi}^{\alpha}_{p_n})^\dagger \sigma^{\eta}_{\alpha,\beta} \hat{\psi}^{\beta}_{p_n} /2$ ($\eta=X,Y,Z$), where $\sigma^{\eta}_{\alpha,\beta}$ are the Pauli matrices and $\hat{\psi}^{\alpha}_{p_n}$ is the annihilation operator of an atom (right panel of Fig.~\ref{fig:scheme}(b)), with momentum state $\ket{\alpha_{p}}_n$.  The Hamiltonian then reduces to:
\begin{equation}
\hat{H}=\hat{H}_{\mathrm{ex}} + \hbar \omega_Z \hat{S}_Z+\hat{H}_{\mathrm{in}}, \quad \hat{H}_{\rm ex}/\hbar=\chi\hat{S}_{+}\hat{S}_{-} \label{eq:h}
\end{equation} 
with the collective spin observables defined as $\hat{S}_{\eta}=\sum_{r=1}^N \hat{s}_{p_r}^{\eta}$.
The first term describes the exchange interaction $\chi\hat{S}_{+}\hat{S}_{-} =  \chi \left(\hat{S}^2 - \hat{S}^2_Z + \hat{S}_Z \right)$ and the second term accounts for the average kinetic energy difference between the momentum states $p_0 + p_r \pm \hbar k$ and $\omega_Z=2\hbar k p_0 /M$ (left panel of Fig.~\ref{fig:scheme}(b)).
The third term $\hat{H}_{\mathrm{in}} = \hat{H}_{\mathrm{in}}^d + \hat{H}_{\mathrm{in}}^c$  has two contributions.  $\hat{H}_{\mathrm{in}}^d /\hbar = \sum_{r=1}^N \frac{2 k p_r}{M} \hat{s}_{p_r}^{Z}$, is the pseudo-spin dependent single particle  kinetic energy. It induces a differential phase for each momentum pseudo-spin, and generates a net spatial separation between the wave packets, each one moving with momentum $\pm \hbar k$, resulting in loss of coherence. 
$\hat{H}_{\mathrm{in}}^c /\hbar = \sum_{r=1}^N \frac{p_r^{2}}{2M \hbar} \hat{I}_{p_r}$, is 
the spin-independent single-particle dispersion. It imparts a global phase that is identical for both $\ket{\Downarrow_{p_r}}$ and $\ket{\Uparrow_{p_r}}$ states but varies with $p_r$,  and causes the position space wave packets to spread with time.  $\hat{I}_{p_r}=(\hat{\psi}_{\Uparrow}^{p_r})^\dagger \hat{\psi}_{\Uparrow}^{p_r} + (\hat{\psi}_{\Downarrow}^{p_r})^\dagger \hat{\psi}_{\Downarrow}^{p_r}$ is the momentum number operator and a constant of motion in our system. While the spin coherence can be restored by the application of a spin-echo pulse in the middle of the evolution, 
the ballistic expansion of each of the wave packets in position space induced by $\hat{H}_{\rm{in}}^c$ cannot be restored.

The exchange interaction term does not commute with $\hat{H}_{\rm{in}}$, and can be used to suppress both wave-packet separation and broadening thanks to the term $\hat{S}^2$ in $\hat{H}_{\rm{ex}}$, which creates a many-body gap, $\chi N$ (inset of Fig.~\ref{fig:scheme}(c)), between the fully symmetric Dicke manifold ($S=N/2$) and the lower symmetry states ($S=N/2-1$)~\cite{rey2008many,norcia2018cavity,smale2019observation,deutsch2010spin,luo2023cavity,niu2025many}. The gap favors spin alignment and acts like a spring connecting the two wave packets, preventing their separation while simultaneously extending the system's coherence~\cite{luo2023cavity}. %regardless of the sign of $\chi$.

%%%%%%%%%%%%%%%%%%%%%%%%%%%%%%%%%%%%%%%%%%%%%%%%%%%%%%%%%%%%%%%%%%%%%%%%%%
%%%%%%%%%%%%%%%%%%%%%%%%%%%%%%%%%%%%%%%%%%%%%%%%%%%%%%%%%%%%%%%%%%%%%%%%%%

%Now we analyze how the Hamiltonian in Eq.~\eqref{eq:h} affects the atomic motion. 
%by effectively modifying the parabola kinetic dispersion for the free particle (left panel of Fig.~\ref{fig:scheme}(c)). 

\textit{Mean-field dynamics.--} 
%%%%%%%%%%%%%%%%%%%%%%%%%%%%%%%%%%%%%%%%%%%%%%%%%%%%%%%%%%%%%%%%%%%%%%%%%%
\begin{figure*}
\includegraphics[width=1.9\columnwidth]{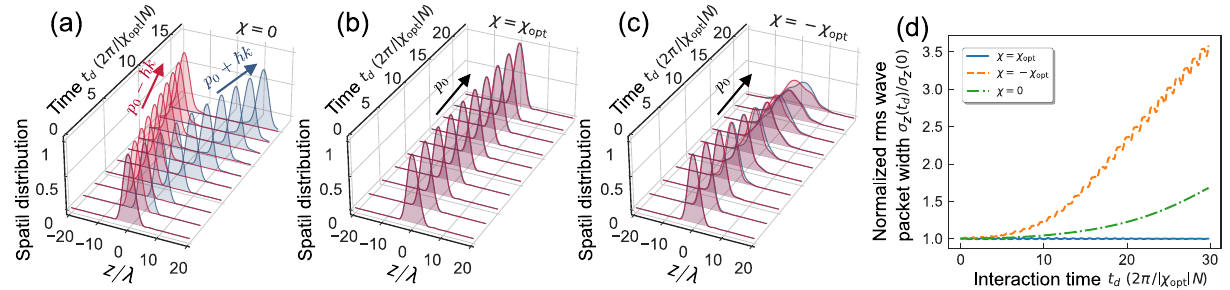}
\caption{\label{fig:soliton} Mean-field simulation of the soliton dynamics, for a given experimental realization sampled from the Wigner function. We show the separate evolution of the density distribution of each wave packet, ignoring interference terms.
After a $\pi/2$ Bragg pulse, the  $p_0 - \hbar k$ (red) and $p_0 + \hbar k$ (blue) wave packets spatially propagate according to  three scenarios:
(a) if $\chi=0$, the wave packets spatially separate while broadening over time as free particles.
(b) if  $\chi = \chi_{\rm{opt}}$, the wave packets merge and maintain their shape, demonstrating soliton behavior.
(c)  if $\chi = -\chi_{\rm{opt}}$, the wave packets merge but show enhanced broadening compared to the free evolution case. (d) Time evolution of the rms width of the spatial profile (using a Gaussian fit), normalized by the initial width. For $\chi=\chi_{\rm opt}$, the broadening is suppressed, consistent with the formation of a soliton.}
%Inset: The interferometer contrast over time $t_d$ remains stable for $\chi = \pm \chi_{\rm opt}$ but decays in the free evolution case due to wave packet separation.}
\end{figure*}
%%%%%%%%%%%%%%%%%%%%%%%%%%%%%%%%%%%%%%%%%%%%%%%%%%%%%%%%%%%%%%%%%%%%%%%%%%
The suppression of the wave-packet broadening can be illustrated for an initial state where all atoms are aligned along the $X$-direction ($\theta=\pi/2$).  Under the mean-field approximation, the Hamiltonian, in a frame rotating at $\omega_Z$, reduces to $\hat{H}_{\rm MF} = \hbar \sum_r {\vec B}_{p_r}\cdot {\vec {\hat{s}}}_{p_r} + \hat{H}^c_{\rm in}$ with $\vec{B}_{p_r}=\{ \chi N,0, \frac{2 k p_r}{M}\}$. In the regime where $|\chi| N \gg 2 k \sigma_{p} /M$ (we set $\chi<0$), the relevant eigenvalues take the form:
% \begin{equation}
% \small
% E_{p_r} \approx - \frac{1}{2} \sqrt{(\chi N)^2 + \left(\frac{2 k p_r}{M} \right)^2} +\frac{p_r^2}{2M \hbar}\approx  \frac{\chi N}{2}  +\Big(1+\frac{4E_R}{\chi N}\Big)\frac{p_r^2}{2M \hbar}, \label{eq:Ep}
% \end{equation} 
\begin{equation}
E_{p_r} \approx - \frac{1}{2} |\vec{B}_{p_r}| +\frac{p_r^2}{2M \hbar}\approx  \frac{\chi N}{2}  +\Big(1+\frac{4E_R}{\chi N}\Big)\frac{p_r^2}{2M \hbar}, \label{eq:Ep}
\end{equation} 
where $E_R=\hbar k^2/2M$ is the recoil energy.
%The corresponding eigenstates can be approximated as $\ket{\pm x}$, with the lower eigenenergy branch shown in Fig.~\ref{fig:scheme}(b) (purple arrows). 
Note that such a type of dispersion emulates the dynamics of a relativistic particle, and has been realized with a single trapped ion~\cite{gerritsma2010quantum} or in BECs subject to spin-orbit coupling~\cite{leblanc2013direct}.
In our case, Eq.~\eqref{eq:Ep} describes a particle with a rest mass energy $M^*= M/(1+ 4E_R/\chi N)$ and an effective speed of light $c_s=\sqrt{N|\chi /2M^*|}$. When 
\begin{equation}
N \chi_{\rm opt} =  -4E_R,
\end{equation} $M^*\to \infty$ and results in a flattened dispersion (right panel of Fig.~\ref{fig:scheme}(d)). Consequently, at $\chi_{\rm opt}$ the wave packets propagate as a soliton, features no dispersion. We emphasize that similar flattening of the band can be realized for non-interacting atoms with an additional driving field along the $X$-axis with Rabi frequency $\Omega=4E_R$. However,  due to the U(1) symmetry of the exchange interaction, a soliton  stabilized by the self-generated field 
can exist regardless of the initial orientation of the collective spin in the $X-Y$ plane (see Fig.~\ref{fig:scheme}(c) for arbitrary initial state on the equator); As a result, the protection will not be canceled by a spin echo pulse.

\begin{figure}
\includegraphics[width=1.\columnwidth]{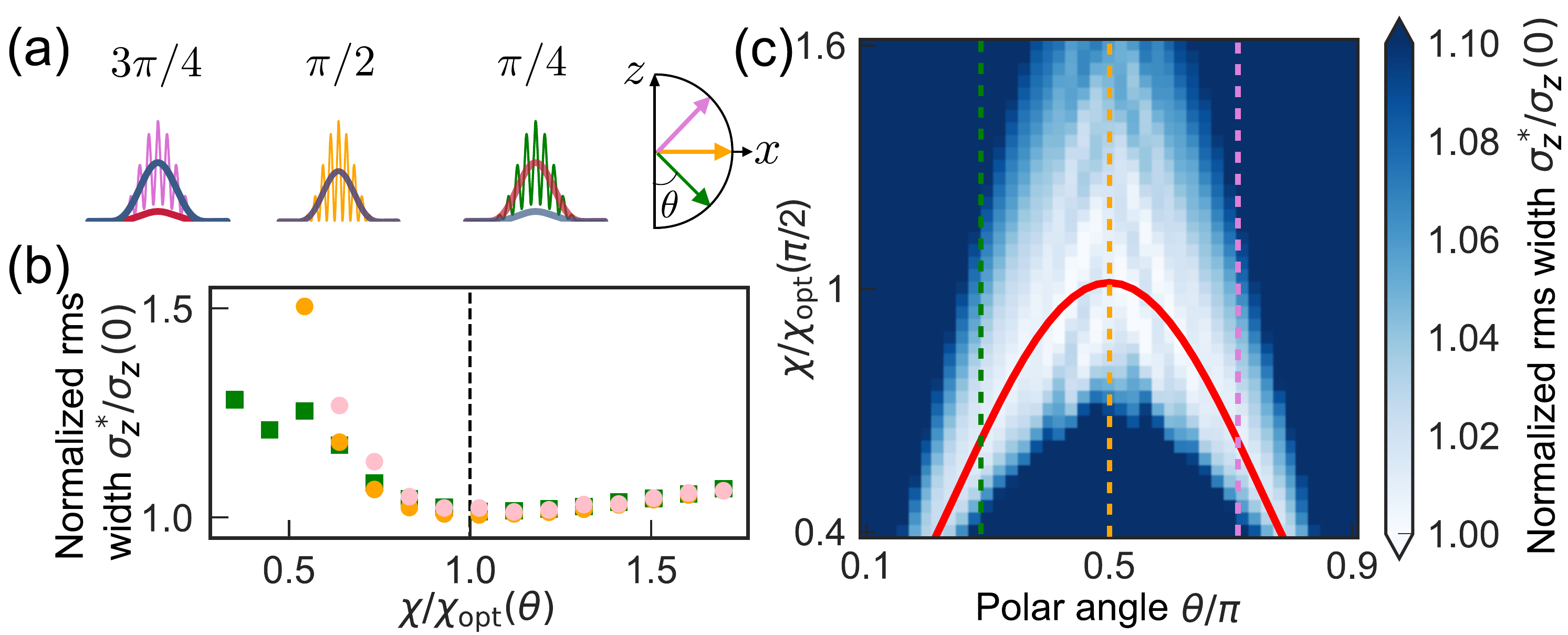}
\caption{\label{fig:theta} Wave-packet dynamics for arbitrary polar angle $\theta$ using mean-field simulations. 
(a) Schematic of the wave packets in position space for initial polar angle $\theta=\pi/4$ (green), $\pi/2$ (orange) and $3\pi/4$ (pink). (b) Normalized rms width $\sigma_z^*$ vs interaction strength $\chi$  at $t_d= 2\pi \times  30 /(|\chi_{\rm opt}| N)$. The  dashed lines indicate $\chi=\chi_{\rm opt}(\pi/2)$. (c) Normalized rms width $\sigma_z^*$ vs $\theta$ and $\chi$ at  $t_d$.  The red curve is the optimal interaction strength as given by Eq.~\eqref{eq:theta}. The cuts for  $\theta=\pi/4,\pi/2,3\pi/4$  are shown  in panel (b).} 
\end{figure}

While so far we focused on  $\theta=\pi/2$, solitons can also be engineered for any  $\theta$, as shown in Fig.~\ref{fig:theta}(a). The only difference is that in this case, there is an additional Ising term in the exchange interactions $ -\chi \hat{S}^2_Z$, which at the mean-field level changes $\vec{B}_{p_r}\to\{ \chi N,0, \frac{2 k p_r}{M}  -\chi N \cos\theta \}$. The latter generates a net precession that needs to be accounted for. As detailed in the SM~\footnote{See Supplemental Material at [URL will be inserted by publisher] for details of ..., includes Ref.} the optimal interaction strength is given by:
\begin{equation}
\chi_{\rm opt}(\theta) = \chi_{\rm opt} \sin^2\theta
\label{eq:theta}
\end{equation}  with a modified average momentum $p_0 - \hbar k \cos\theta$.
Notably, the optimal interaction strength decreases as the initial state approaches either the north or south poles. However, the condition $|\chi_{\rm opt}(\theta)| N \gg 2k \sigma_p/M$ needs to be enforced at all time, a requirement that significantly reduces  the allowed $\sigma_p$ in the vicinity of the poles.

The wave-packet dynamics can be simulated using the mean-field equations of motion~\cite{Note1}, with results shown in Fig.~\ref{fig:soliton}. We only show a single realization  to facilitate visualization. The thermal average needs to be  realized by sampling over the Wigner function. 
For a free particle with $\chi=0$ (Fig.~\ref{fig:soliton}(a)), the two sampled wave packets separate and broaden over time. In contrast, the wave packets remain bound to each other for $\chi=\pm \chi_{\rm opt}$ as shown in Fig.~\ref{fig:soliton}(b) and (c). 
At the optimal interaction strength $\chi=\chi_{\rm opt}$ (Fig.~\ref{fig:soliton}(b)), the wave packets maintain their initial shape during propagation, generating a  soliton.  
On the other hand, for $\chi=-\chi_{\rm opt}$ (Fig.~\ref{fig:soliton}(c)), the wave packets not only broaden over time but even deviate from their Gaussian profiles at later times.

The evolution of the  wave packets is quantitatively analyzed in Fig.~\ref{fig:soliton}(d), where we track the rms width in position space $\sigma_z(t_d)$ by fitting the wave packets as Gaussian functions over time $t_d$. The ratio of the rms width to its initial value is plotted as a function of time. For $\chi=\chi_{\rm opt}$, the rms width remains constant over an extended period, when the soliton is formed. In contrast, the free evolution case exhibits typical diffusion dynamics, while for $\chi=-\chi_{\rm opt}$, the wave packets broaden more than the free evolution case. 
%%%%%%%%%%%%%%%%%%%%%%%%%%%%%%%%%%%%%%%%%%%%%%%%%%%%%%%%%%%%%%%%%%%%%%%%%%
%%%%%%%%%%%%%%%%%%%%%%%%%%%%%%%%%%%%%%%%%%%%%%%%%%%%%%%%%%%%%%%%%%%%%%%%%%

Fig.~\ref{fig:theta}(b) shows the ratio between the fitted rms width $\sigma_z^*$ after a long evolution time, versus the initial width, for different initial polar angles $\theta$. The solutions confirm that the optimal exchange interaction strength follows Eq.~\eqref{eq:theta} (black dashed line). Furthermore, Fig.~\ref{fig:theta}(c) reveals that the soliton is robust and persists over a broad range of interaction strengths around the predicted optimal value (red solid line). The robustness is stronger close to the equator where $\chi_{\rm opt}(\theta)$ reaches a maximum.  

%%%%%%%%%%%%%%%%%%%%%%%%%%%%%%%%%%%%%%%%%%%%%%%%%%%%%%%%%%%%%%%%%%%%%%%%%%
\begin{figure}
\includegraphics[width=1.\columnwidth]{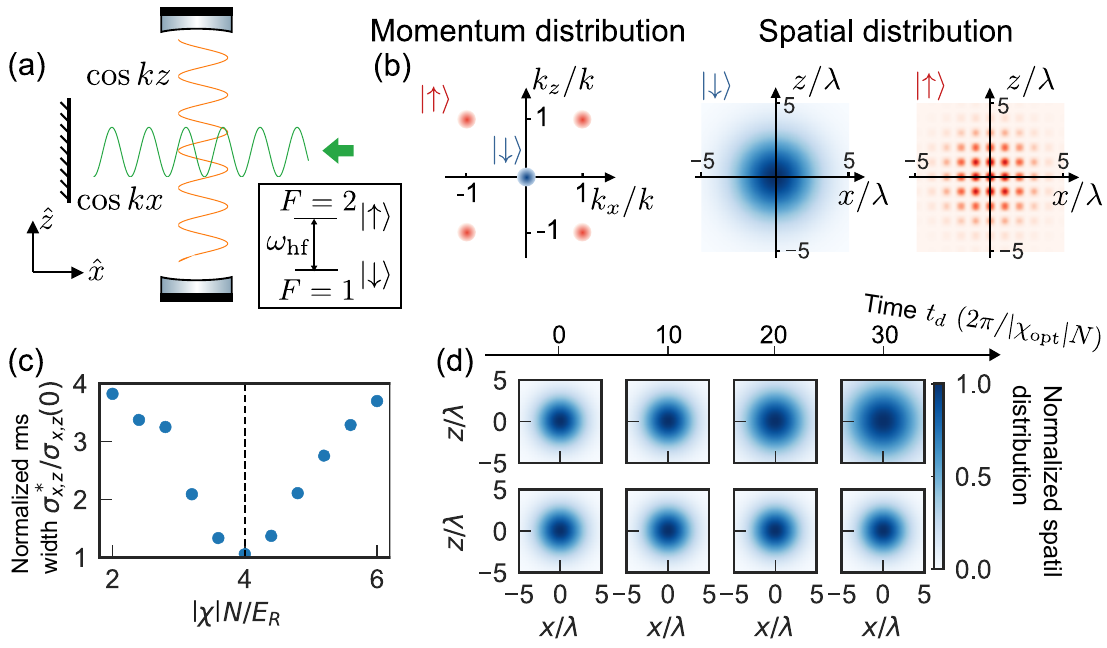}
\caption{\label{fig:2d} Proposed scheme to stabilize a soliton in 2D. (a) It uses a standing-wave cavity along the
$z$-direction (orange) and a retro-reflected drive field along the $x$-direction (green). (b) The initial state is prepared via a Raman pulse as a superposition between $\ket{0,0;\downarrow}$ and $\ket{\pm \hbar k,\pm \hbar k;\uparrow}$ in the 2D momentum-spin coupled basis. We plot the initial momentum and spatial distributions for the internal states $\ket{\downarrow}$ and $\ket{\uparrow}$.
(c) Normalized rms width $\sigma_{x,z}^*$ for $\ket{\downarrow}$ vs $\chi$ at $t_d = 2\pi \times 100/(|\chi_{\rm opt}| N)$, with the optimal interaction strength  given by $\chi_{\rm opt}=-4E_R$.
(d) Simulated spatial distribution vs time for $\ket{\downarrow}$. Top: $\chi=0$, the wave packets broaden over time.
Bottom: $\chi=\chi_{\rm opt}$, the wave packets retain their initial shape without broadening.
}
% (c) Simulated position space density for $\ket{\uparrow}$ at $t_d = 2\pi \times 12/(|\chi_{\rm opt}| N)$. 
% Left: $\chi = 0$, the initial interference pattern (red) gradually evolves, leading to the separation of different momentum states in position space at $t_d$ (blue).
% Middle: $\chi = \chi_{\rm opt}$, the wave packets merge and maintain the initial distribution.
% Right: $\chi =-\chi_{\rm opt}$, the wave packets merge but exhibit significant broadening. Here $n_{x,y}$ is the normalized density in position space with respect to the peak density at a given time.} 
\end{figure}
%%%%%%%%%%%%%%%%%%%%%%%%%%%%%%%%%%%%%%%%%%%%%%%%%%%%%%%%%%%%%%%%%%%%%%%%%%

\textit{Solitons in 2D and 3D.--}The above mechanism can be generalized to a 2D system by adding a drive perpendicular to the cavity axis,  as shown in Fig.~\ref{fig:2d}(a). 
In this case, one needs to first start with atoms in internal state $\ket{\downarrow}$, and with momentum centered around zero  $\vec{p}_0=(0,0)$   by velocity selection. The momentum being zero relative to the frame defined by the optical cavity is important since it enables resonant Raman coupling to four momentum states  $(\pm \hbar k,\pm \hbar k)$ all separated by  $\omega_Z=2 E_R+ \omega_{\rm hf}$ from $(0,0)$.  Here  $\omega_{\rm hf}$ is the energy splitting between  $\ket{\downarrow}$ and another  hyperfine level $\ket{\uparrow}$, i.e. between the $F=2$ and $F=1$ ground  hyperfine levels in  $^{87}\mathrm{Rb}$ atoms (inset of Fig.~\ref{fig:2d}(a)).   Note that we added an additional internal degree of freedom to increase the energy difference, $\omega_z$, by $\omega_{\rm hf}$  since in this case it is not boosted by a finite $p_0$ as was the case in the 1D case.

After velocity selection, a follow-up 
 Raman $\pi/2$ pulse,  can prepare an atom $r$ in the array to the superposition state 
 $\frac{1}{\sqrt{2}}(\ket{\Downarrow_{\vec{p}}}_r + \ket{\Uparrow_{\vec{p}}}_r)$ with 
$\ket{\Downarrow_{\vec{p}=(p_x,p_z)}}_r \equiv \ket{p_x,p_z,\downarrow}_r$, 
and $\ket{\Uparrow_{\vec{p}}}_r \equiv \frac{1}{2}\sum_{\mu_x,\mu_z=\pm1}\ket{p_x + \mu_x k, p_z + \mu_z k,\uparrow}_r$, with the initial momentum and spatial distribution shown in Fig.~\ref{fig:2d}(b).

Time evolution in the driven cavity enable processes in which an atom in $\ket{\Downarrow_{\vec{p}}}$ absorbs a pump field photon (green),  subsequently emitting a cavity photon (orange) while flipping the spin to $\ket{\Uparrow_{\vec{p}}}$. The emitted photon is then absorbed by another $\ket{\Uparrow_{\vec{q}}}$ atom, which then emits a green pump photon while flipping to $\ket{\Downarrow_{\vec{q}}}$. As in the 1D case, these processes generate exchange interaction within the aforementioned spin-1/2 system. A similar type of processes have demonstrated in recent experiments~\cite{baumann2010dicke,Mivehvar2021}.

In this situation again, one finds a modified energy spectrum similar to Eq.~\eqref{eq:Ep}, but instead of $(2 k p_r/M)^2$ one has,  $\sum_{\mu_x,\mu_z=\pm1}[(k p_x/M)^2+ (k p_z/M)^2+2(k \mu_x p_x/M) (k \mu_z p_z/M)] =(2 k p_x/M)^2+ (2 k p_z/M)^2$. Therefore, again  by choosing $ N\chi_{\rm opt}=-4E_R$, one can cancel the dispersion but now in two directions~\cite{Note1}. 

In Fig.~\ref{fig:2d}(c), we fit the position space density of $\ket{\downarrow}$ internal state using a 2D Gaussian function and plot the ratio of the rms width at $t_d=2\pi\times 100 / (|\chi_{\rm opt}| N)$, $\sigma_{x,z}^*$, to its initial width as a function of the interaction strength. 
In Fig.~\ref{fig:2d}(d), we show the position-space density of the internal state $\ket{\downarrow}$ at different interaction times. For the non-interacting case ($\chi=0$, top panel), the density profile broadens over time, whereas for the optimal interaction strength ($\chi=\chi_{\rm opt}$, bottom panel), the initial shape is preserved. % Due to the presence of four peaks in the momentum space, the initial state exhibits an interference pattern (red density plot, left panel). Without exchange interactions ($\chi = 0$), the wave packets separate in position space over time (blue density plot, left panel). In contrast, for $\chi = \pm \chi_{\rm opt}$, the wave packets remain bounded, preserving an interference pattern (middle and right panels). At the optimal interaction strength $\chi = \chi_{\rm opt}$, the wave packets maintain their initial shape, whereas for $\chi = -\chi_{\rm opt}$, they broaden significantly.
Interestingly, compared to the well-known self-organization transition~\cite{baumann2010dicke,Mivehvar2021}, the exchange interaction preserves the $Z_2$ checkerboard pattern even when the transverse drive is weak.

A 3D soliton can be engineered in a similar way but by replacing the transverse retroreflected beam  along $\hat{x}$, i.e., perpendicular to the cavity by two orthogonal retroreflected beams now propagating along $ \hat{x}\pm \hat{y}$ directions, both again perpendicular to the cavity axis (see END MATTER). 

\textit{Experimental considerations.--} 
In contrast to the narrow momentum distribution featured by a BEC, which can be imaged through time-of-flight measurements, a thermal gas presents the challenge of a much broader initial profile even after velocity selection~\cite{luo2023cavity}. This makes it difficult to experimentally observe the suppressed dispersion, as atoms will hit the lower cavity mirror before they expand for sufficient observation time. 
For then 1D soliton we  propose a Raman interferometry scheme amenable to probe the emergent soliton in a thermal gas system with the details in~\cite{Note1}.
The key idea is to use an additional external drive, instead of the exchange interactions, to prepare a soliton between a pair of momentum states in another hyperfine level $\ket{0}$, and inject it into one arm of the interferometer. 
We choose the cavity frequency to be far from resonant from the atomic transition for this internal state, to make it insensitive to exchange interactions. In the other arm, we inject a soliton between another pair of momentum states in a hyperfine state $\ket{1}$ that experiences exchange interactions. The contrast achieved at the output of the interferometer serves as a probe of the differential dispersion. 
This spectroscopic method could be also used to detect 2D  and 3D solitons. In this case, however,  it would be required to have access to four different internal levels, so the use of a BEC would make the detection easier. 

Decoherence sources such as free space scattering and superradiance from photons leaking out of the cavity, not included so far,  can disrupt the coherence between momentum states and destroy the soliton ~\cite{luo2023cavity},  with the latter being the most relevant in large arrays. One way to mitigate superradiance, as discussed in the SM~\cite{Note1}, is the use of a dual pump tone configuration~\cite{luo2025hamiltonian,luo2024realization}. 

\textit{Conclusion and outlook.--}
We have proposed a scheme to manipulate atomic motion by leveraging the interplay between single-particle dispersion and many-body cavity interactions and came up with a new mechanism for generating solitons in 1D, 2D  and 3D geometries without the need for quantum degeneracy. For thermal atoms, we proposed an interferometric detection protocol that can be used to spectroscopically resolve the engineered dispersion~\cite{balents2020superconductivity}. 
Our work opens many different opportunities. In the context of quantum simulation, while we have neglected contact interactions, in their presence, it could enable investigations on their enhanced role in a flat band system. For quantum sensing,  our system could be an interesting resource towards the path of miniaturization on the one hand, and on the other hand towards inertial sensing in outer space via the solitons in 2D and 3D. The solitons should enable longer interrogation times, favorable for tests of the weak equivalence principle in free fall atoms, for instance,  by detecting differential acceleration between isotopes. Finally, while most of the discussion was restricted to a mean-field analysis, beyond mean-field effects can in addition generate spin squeezing for quantum-enhanced interferometry~\cite{greve2022entanglement,wilson2024entangled}. 

{\it Acknowledgments}
We thank Chitose Maruko and  Yang Yang for useful feedback on the manuscript.  This material is based upon work supported by   the Vannevar Bush Faculty Fellowship, the Heising-Simons foundations, the NSF JILA-PFC PHY-2317149 and OMA-2016244 (QLCI), the U.S. Department of Energy, Office of Science, National Quantum Information Science Research Centers, Quantum Systems Accelerator and NIST. 

\bibliography{reference}

\begin{widetext}
\begin{center}
\textbf{END MATTER}
\end{center}
\end{widetext}

\textit{Solitons in higher dimensions.--} 
In this section, we discuss a protocol for generating two-dimensional solitons. It uses concepts similar to those discussed previously, but with the necessary modifications to address the challenges of higher-dimensional systems.
To notice  such challenges, imagine   we couple  two momentum states $(p_x,p_z)$ and $(p_x+\hbar k_x,p_z+\hbar k_z)$ through the cavity by exchanging photons  with a  wave vector $(k_x,k_z)$, the resulting effective dispersion becomes:
\begin{widetext}
\begin{equation}
\begin{aligned}
E_{\vec{p}=(p_x,p_z)}&=-\frac{1}{2}\sqrt{(\chi N)^2+\Big (\frac{ k_{x} p_x +  k_z p_z}{M}\Big)^2} + \frac{p_x^2 + p_z^2}{2M\hbar} 
\approx \frac{\chi N}{2} + \frac{p_x^2}{2M\hbar} \Big(1 + \frac{\hbar k^2_x}{2M \chi N}\Big) 
+ \frac{p_z^2}{2M\hbar} \Big(1 + \frac{\hbar k^2_y}{2 M \chi N}\Big) + \frac{p_x p_z k_x k_z}{2M^2 \chi N},
\end{aligned}
\end{equation}
\end{widetext}
The cross term proportional to $p_x p_z$ cannot be canceled, resulting in an  anisotropic dispersion that broadens the wave packets instead of a soliton.

We  propose an alternative scheme to form a 2D soliton via the superposition of five momentum states centered around $(0,0)$ and $(\pm \hbar k,\pm \hbar k)$. The main idea behind is  the use on additional retroreflected beam perpendicular to the cavity direction,  that creates a standing wave spatial profile $\propto \cos(k x)\cos(k y)$ as discussed in the main text,  and  an additional internal state degree of freedom, labeled $\ket{\uparrow}$ and $\ket{\downarrow}$. The internal levels are  separated by an energy splitting $\omega_{\rm hf}$. 
This setting closely related to the well-known self-organization transition broadly explored in cold gases~\cite{baumann2010dicke,Mivehvar2021}.
Following Ref.~\cite{Mivehvar2021} the effective Hamiltonian in the far detunned limit corresponds to an LMG model: $\hat{H}_{\rm LMG}+\hat{H}_{\rm dis}$ with $\hat{H}_{\rm LMG}=\omega_Z \hat{S}_Z+\omega_Z \hat{S}_Z +\chi \hat{S}^2_X $  when written in terms of  the spin-momentum states   $\ket{\Downarrow_{\vec{p}}} \equiv \ket{\vec{p},\downarrow}_r$ and $
\ket{\Uparrow_{\vec{p}}}_r = \frac{1}{2}\sum_{\mu_x,\mu_z=\pm}\ket{\Uparrow_{\vec{p},\mu_x,\mu_z}}_r$,   with $\ket{\Uparrow_{\vec{p},\mu_x,\mu_z}}_r \equiv \ket{p_x+\mu_x \hbar k, p_z+\mu_z \hbar k,\uparrow}_r$.

The single-particle kinetic Hamiltonian,  reads:
\begin{widetext}
\begin{equation}
\begin{aligned}
\hat{H}_{\rm dis} &= \sum_{r}\frac{p_{x,r}^2+p_{z,r}^2}{2M}\hat{n}_{\vec{p}_r}^{\Downarrow} +\sum_{r,\mu_x,\mu_z}\left(\omega_{\rm hf}+ \frac{(p_{x,r}+\mu_x\hbar k)^2+(p_{z,r}+\mu_z\hbar k)^2}{2M}\right)\hat{n}_{\vec{p}_r,\mu_x,\mu_z}^{\Uparrow} \\
&=\sum_{r}\frac{p_{x,r}^2+p_{z,r}^2}{2M}\hat{I}_{\vec{p}_r} +\sum_{r,\mu_x,\mu_z}\left(\omega_Z+\frac{\mu_x \hbar k p_{x,r}+\mu_z \hbar k p_{z,r}}{M}\right)\hat{n}_{\vec{p}_r,\mu_x,\mu_z}^{\Uparrow},
\end{aligned}
\end{equation}
\end{widetext}
with  $\ket{\Uparrow_{\vec{p}},\mu_x,\mu_z}_r \equiv (\hat{\psi}_{\vec{p}_r}^{\Uparrow})^\dagger\ket{0}$ and $\ket{\Downarrow_{\vec{p}}}_r \equiv (\hat{\psi}_{\vec{p}_r}^{\Downarrow})^\dagger\ket{0}$ and 
$\hat{n}_{\vec{p}_r}^{\Downarrow}=(\hat{\psi}_{\vec{p}_r}^{\Downarrow})^\dagger\hat{\psi}_{\vec{p}_r}^{\Downarrow}$ and $\hat{n}_{\vec{p}_r,\mu_x,\mu_z}^{\Uparrow}=(\hat{\psi}_{\vec{p}_r,\mu_x,\mu_z}^{\Uparrow})^\dagger\hat{\psi}_{\vec{p}_r,\mu_x,\mu_z}^{\Uparrow}$. The identity operator is defined as $
\hat{I}_{\vec{p}_r}=\hat{n}_{\vec{p}_r}^{\Downarrow}+\sum_{\mu_x,\mu_z=\pm}\hat{n}_{\vec{p}_r,\mu_x,\mu_z}^{\Uparrow}$. Notice that above we take the large pump-cavity detuning and weak pump field limit to ignore the extra lattice trapping potential due to the ac-stark shift from either cavity or pump field~\cite{baumann2010dicke}.

To generate the desired exchange interactions one needs  $\chi N\ll \omega_Z$. 
Unlike the 1D case, here the momentum boost $p_0$ is absent, significantly reducing the average kinetic energy difference to $ 2E_R$.
But the additional internal state energy difference can come to the rescue (this is why we needed another internal state)  making the average energy difference between the states $\omega_Z =\omega_{\rm hf}+2E_R$. If  $\chi N\ll \omega_Z$     terms that does not preserve the total magnetization are rotated out, thus restoring the interacting part of the Hamiltonian to an exchange interacting model.    

\begin{equation}
\begin{aligned}
\hat{H}_{\rm ex} &= \chi \sum_{n,m}(\hat{\psi}_{\vec{p}_n}^{\Uparrow})^\dagger \hat{\psi}_{\vec{p}_n}^{\Downarrow}(\hat{\psi}_{\vec{p}_m}^{\Downarrow})^\dagger \hat{\psi}_{\vec{p}_m}^{\Uparrow} \\
&\approx \frac{\chi N}{2}\sum_n\left[(\hat{\psi}_{\vec{p}_n}^{\Uparrow})^\dagger\hat{\psi}_{\vec{p}_n}^{\Downarrow}+(\hat{\psi}_{\vec{p}_n}^{\Downarrow})^\dagger\hat{\psi}_{\vec{p}_n}^{\Uparrow}\right],
\end{aligned}
\end{equation}

 %We denote the corresponding annihilation operators for the particle at these states as $\hat{\psi}_{\vec{p}_r}^\Downarrow$ and $\hat{\psi}_{\vec{p}_r,\mu_x,\mu_z}^\Uparrow$.

% %%%%%%%%%%%%%%%%%%%%%%%%%%%%%%%%%%%%%%%%%%%%%%%%%%%%%%%%%%%%%%%%%%%%%%%%%%
\begin{figure}
\includegraphics[width=1\columnwidth]{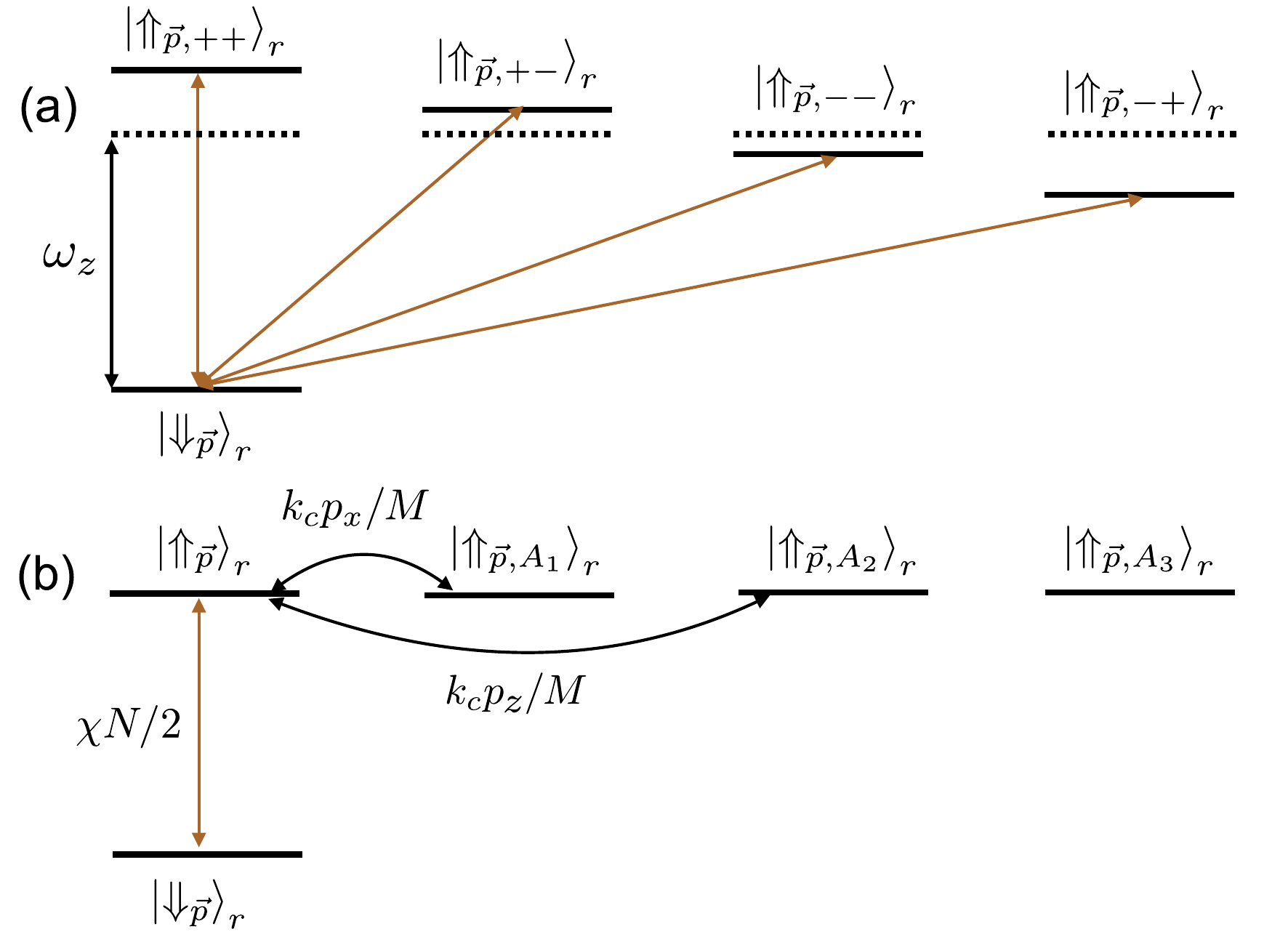}
\caption{\label{fig:couple2d} Mean-field coupling diagram in 2D for atom $r$. (a) The system consists of five states in the bare basis: $\ket{\Downarrow_{\vec{p}}} \equiv \ket{\vec{p},\downarrow}_r$ and four  momentum states $\ket{\Uparrow_{\vec{p},\mu_x\mu_z}}_r \equiv \ket{p_x+\mu_x \hbar k, p_z+\mu_z \hbar k,\uparrow}_r$ with $\mu_x, \mu_z = \pm$. The single-particle energy difference between them is given by $\omega_Z + (p_x \mu_x k + p_z \mu_z k)/M$. The exchange interaction (orange arrows) couples $\ket{\Downarrow_{\vec{p}}}_r$ to all  four  momentum states $\ket{\Uparrow_{\vec{p},\mu_x\mu_z}}_r$. (b)  Since the exchange interaction only couples $\ket{\Downarrow_{\vec{p}}}$ to the symmetric superposition of the four momentum states $\ket{\Uparrow_{\vec{p}}}_r$, it is convenient  to use  a  different basis, defined in Eq.~\eqref{eq:dress}. In this basis, the   single-particle inhomogeneity is no longer diagonal  and has couplings between the states, $\ket{\Uparrow_{\vec{p},A_1}}_r$ and $\ket{\Uparrow_{\vec{p},A_2}}_r$, with strengths $k p_x / M$ and $k p_z / M$, respectively.
}
\end{figure}
% %%%%%%%%%%%%%%%%%%%%%%%%%%%%%%%%%%%%%%%%%%%%%%%%%%%%%%%%%%%%%%%%%%%%%%%%%%
For convenience, we define another complete set of orthogonal states, starting from the symmetric superposition $\ket{\Uparrow_{\vec{p}}}_r$. The other three orthogonal dressed states are given by:
\begin{equation}
\begin{aligned}
\ket{\Uparrow_{\vec{p},A_1}}_r&= \frac{1}{2}\left(\ket{\Uparrow_{\vec{p},+,+}}_r+\ket{\Uparrow_{\vec{p},+,-}}_r-\ket{\Uparrow_{\vec{p},-,+}}_r-\ket{\Uparrow_{\vec{p},-,-}}_r\right)\\
\ket{\Uparrow_{\vec{p},A_2}}_r&= \frac{1}{2}\left(\ket{\Uparrow_{\vec{p},+,+}}_r-\ket{\Uparrow_{\vec{p},+,-}}_r+\ket{\Uparrow_{\vec{p},-,+}}_r-\ket{\Uparrow_{\vec{p},-,-}}_r\right)\\
\ket{\Uparrow_{\vec{p},A_3}}_r&=\frac{1}{2}\left(\ket{\Uparrow_{\vec{p},+,+}}_r-\ket{\Uparrow_{\vec{p},+,-}}_r-\ket{\Uparrow_{\vec{p},-,+}}_r+\ket{\Uparrow_{\vec{p},-,-}}_r\right).\label{eq:dress}
\end{aligned}
\end{equation}

In  this basis, the matrix that governs the dynamics of the atoms is  given by (Fig.~\ref{fig:couple2d}(a,b)): $\{\ket{\Downarrow_{\vec{p}}}_r, \ket{\Uparrow_{\vec{p}}}_r, \ket{\Uparrow_{\vec{p},A_1}}_r, \ket{\Uparrow_{\vec{p},A_2}}_r, \ket{\Uparrow_{\vec{p},A_3}}_r\}$ is given by:
\begin{equation}
\left(\begin{array}{ccccc}
0 & \frac{\chi N}{2} & 0 & 0 & 0\\
\frac{\chi N}{2} & 0 & \frac{k_{c}p_{x}}{M} & \frac{k_{c}p_{z}}{M} & 0\\
0 & \frac{k_{c}p_{x}}{M} & 0 & 0 & 0\\
0 & \frac{k_{c}p_{z}}{M} & 0 & 0 & 0\\
0 & 0 & 0 & 0 & 0
\end{array}\right).
\end{equation}

When a  $\pi/2$ Raman  pulse is applied ---generated via a two-photon Raman transition by pumping both the cavity and a transverse field---
$\frac{1}{\sqrt{2}}\left(\ket{\Downarrow_{\vec{p}}}_r+\ket{\Uparrow_{\vec{p}}}_r\right)$ one in fact prepares a state that has  substantial overlap with  an eigenstate  of the matrix above   with an  eigenenergy :
\begin{equation}
\begin{aligned}
E_{\vec{p}} &= - \sqrt{\left(\frac{\chi N}{2}\right)^2 + \left(\frac{k p_x}{M}\right)^2+\left(\frac{k p_z}{M}\right)^2} + \frac{p_x^2+p_z^2}{2M\hbar} \\
 &\approx \frac{\chi N}{2}+\left(1+\frac{4E_R}{\chi N}\right)\frac{p_x^2+p_z^2}{2M\hbar}.
\end{aligned}
\end{equation}
Now, similar to the 1D scenario, by choosing the interaction strength to be  $\chi_{\rm opt} N = -4 E_R$, one can remove the dispersion at the leading order, enabling  the formation of a stable 2D soliton.

\begin{figure}
\includegraphics[width=1\columnwidth]{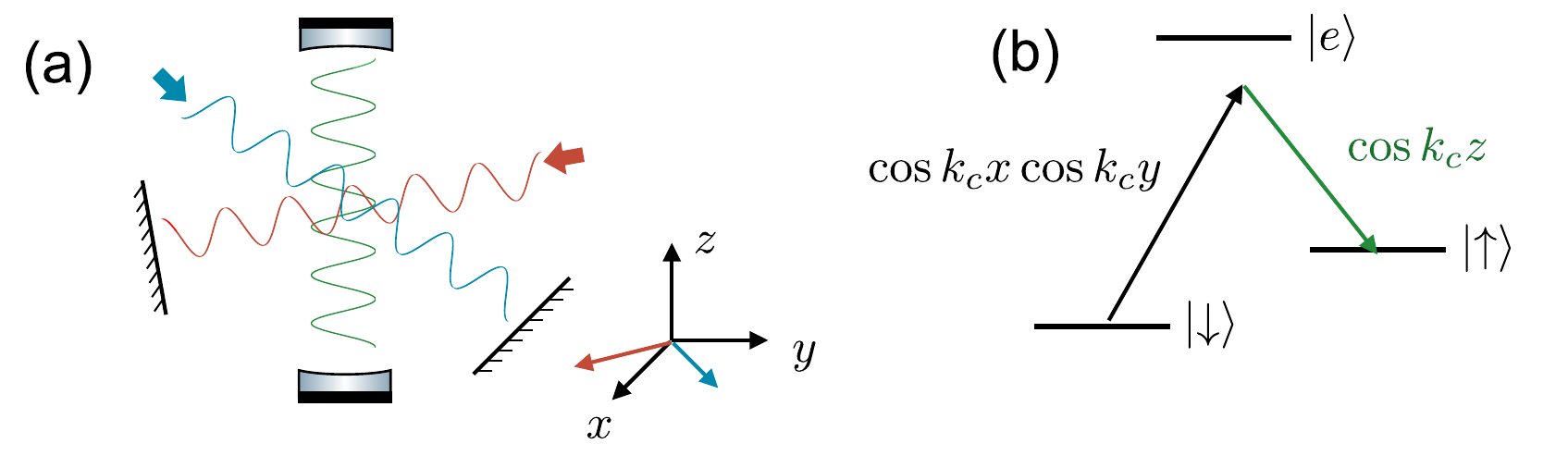}
\caption{\label{fig:3D} (a) Proposed scheme to stabilize a soliton in 3D, using a standing-wave cavity along the
$z$-direction (green) and two retro-reflected drive fields along the $x\pm y$-direction (red and blue). (b) Cavity-assist two-photon transition between $\ket{\uparrow}$ and $\ket{\downarrow}$ internal state with a spatial profile $\propto \cos k x \cos k y \cos k z$.} 
\end{figure}

A 3D soliton can also be stabilized using the setup shown in Fig.~\ref{fig:3D}(a). Here, we have the cavity field along the $z$ (green) direction and two transverse retroreflected beams along the $x\pm y$ directions (blue and red). These two fields create an interference pattern with a spatial profile $\propto \cos (kx+ky)+ \cos (kx-ky) = 2\cos k x \cos k y$.
Using them,  a Raman pulse can be used to prepare, via a two-photon transition an initial state that is a coherent superposition between $\ket{0,0,0,\downarrow}$ and $\ket{\pm \hbar k,\pm \hbar k,\pm \hbar k,\downarrow}$, as shown in Fig.~\ref{fig:3D}(b). In this system the exchange interaction is generated by exchanging photons in the cavity as follows: an atom in $\ket{\Downarrow_{\vec{p}}}$ absorbs a pump field photon with spatial profile $\propto \cos k x \cos k y$, subsequently emitting a cavity photon (green) and flipping the spin to $\ket{\Uparrow_{\vec{p}}}$. The emitted photon is then absorbed by another atom in state $\ket{\Uparrow_{\vec{q}}}$, which emits a pump photon while flipping its state to $\ket{\Downarrow_{\vec{q}}}$. A similar analysis to the one discussed for the 2D case  is also applicable here, resulting in the same optimal interaction strength, $\chi_{\rm opt}N=-4E_R$.

\end{document}

\title{Solitons in arbitrary dimensions stabilized by photon mediated interactions: Supplemental Materials}
\author{Haoqing Zhang}
\affiliation{JILA, NIST and Department of Physics, University of Colorado, Boulder, Colorado 80309, USA}
\affiliation{Center for Theory of Quantum Matter, University of Colorado, Boulder, Colorado 80309, USA}
\author{Anjun Chu}
\affiliation{JILA, NIST and Department of Physics, University of Colorado, Boulder, Colorado 80309, USA}
\affiliation{Center for Theory of Quantum Matter, University of Colorado, Boulder, Colorado 80309, USA}
\author{Chengyi Luo}
\affiliation{JILA, NIST and Department of Physics, University of Colorado, Boulder, Colorado 80309, USA}
\author{James K. Thompson}
\affiliation{JILA, NIST and Department of Physics, University of Colorado, Boulder, Colorado 80309, USA}
\author{Ana Maria Rey}
\affiliation{JILA, NIST and Department of Physics, University of Colorado, Boulder, Colorado 80309, USA}
\affiliation{Center for Theory of Quantum Matter, University of Colorado, Boulder, Colorado 80309, USA}
\date{\today}
\maketitle

\section{System}
Here we provide a detailed derivation of the wave-packet dynamics, in terms of a spin model, following a description recently derived in Ref.~\cite{luo2023cavity}.
The initial state consists of a wave packet  with average momentum $p_0$ and width $\sigma_p$ prepared from an array of thermal atoms  via  velocity selection--- implemented  by a Raman  pulse---.

Omitting other labels irrelevant for the dynamics, we focus on a specific experimental realization with an initial set of position and momentum states labeled as $\{p_1, p_2, \dots p_N\}$, sampled from the Gaussian distribution with width $\sigma_p$.

The preparation step finished with a  two-photon Bragg transition, that prepares a  superposition between $\ket{\Downarrow_p}_n \equiv \ket{p_0 + p}_n$ and $\ket{\Uparrow_p}_n \equiv \ket{p_0 + p + 2\hbar k}_n$, with $k$ being the wave vector of the cavity model and $\sigma_p\ll 2\hbar k$.

\subsection{Spin model}
As explained in the main text, we define the pseudospin-1/2 operators written in  terms of   annihilation operators $\hat{\psi}^\Downarrow_{p_n}$ for $\ket{\Downarrow_p}_n$ momentum state and $\hat{\psi}^\Uparrow_{p_n}$ for $\ket{\Uparrow_p}_n$ momentum state.  
 They are given by 
\begin{equation}
\begin{aligned}
\hat{s}^x_{p_n} &= \frac{1}{2}\left( (\hat{\psi}^\Uparrow_{p_n})^\dagger\hat{\psi}^\Downarrow_{p_n} + (\hat{\psi}^\Downarrow_{p_n})^\dagger\hat{\psi}^\Uparrow_{p_n}\right)  , \\
\hat{s}^y_{p_n} &=\frac{1}{2 i }\left( (\hat{\psi}^\Uparrow_{p_n})^\dagger\hat{\psi}^\Downarrow_{p_n} - (\hat{\psi}^\Downarrow_{p_n})^\dagger\hat{\psi}^\Uparrow_{p_n}\right)  , \\
\hat{s}^Z_{p_n} &=\frac{1}{2} \left((\hat{\psi}_{\Uparrow}^{p})^\dagger \hat{\psi}_{\Uparrow}^{p} - (\hat{\psi}_{\Downarrow}^{p})^\dagger \hat{\psi}_{\Downarrow}^{p}\right),    
\end{aligned}
\end{equation}
In terms of them we define  collective spin operators
\begin{equation}
    \hat{S}_\alpha = \sum_n \hat{s}^\alpha_{p_n} \quad \alpha\in[x,y,z,+,-].
\end{equation}

After  adiabatically eliminating  the excited state and the cavity field we obtain an effective Hamiltonian in the ground state manifold~\cite{luo2023cavity}, evolving only the motional degree of freedom. It is given by:
\begin{equation}
\hat{H}=\hat{H}_{\mathrm{ex}}+\hat{H}_{\mathrm{in}}
\end{equation} 
The exchange interaction  can be written as:
\begin{equation}
    \hat{H}_{\mathrm{ex}}=\chi\hat{S}_{+}\hat{S}_{-} = \chi \left(\hat{S}^2 - \hat{S}^2_z + \hat{S}_Z \right),
\end{equation}
and the single particle dispersion terms as ,
\begin{equation}
\begin{aligned}
        \hat{H}_{\mathrm{in}} /\hbar &= \sum_n \omega_{p_n}^{\Downarrow} (\hat{\psi}^{\Downarrow}_{p_n})^\dagger \hat{\psi}^{\Downarrow}_{p_n} + \omega_{p_n}^{\Uparrow} (\hat{\psi}^{\Uparrow }_{p_n})^\dagger \hat{\psi}^{\Uparrow }_{p_n} \\
        &=\sum_n \left( \omega_Z + \frac{2 k p_n}{M} \right) \hat{s}^Z_{p_n} + \Bigr[\frac{p_n^{2}}{2M \hbar}+\frac{\left(p_{0}+\hbar k_{c}\right)p_n}{M \hbar} \Bigr] \Bigr[ (\hat{\psi}^{\Downarrow}_{p_n})^\dagger \hat{\psi}^{\Downarrow}_{p_n} +(\hat{\psi}^{\Uparrow }_{p_n})^\dagger \hat{\psi}^{\Uparrow }_{p_n} \Bigr] ,
\end{aligned}\label{eq:H}
\end{equation}
Here $M$ is  the atom's mass.
The single-particle energies above are given by
\begin{equation}
    \begin{aligned}
        \hbar \omega_{p}^{\Downarrow}&=\frac{\left(p_{0}+p\right)^{2}}{2M}=\frac{p_{0}^{2}}{2M}+\frac{p_{0}p}{M}+\frac{p^{2}}{2M}\\
        \hbar \omega_{p}^{\Uparrow}&=\frac{\left(p_{0}+2\hbar k_{c}+p\right)^{2}}{2M}=\frac{\left(p_{0}+2\hbar k_{c}\right)^{2}}{2M}+\frac{\left(p_{0}+2\hbar k_{c}\right)p}{M}+\frac{p^{2}}{2M}
    \end{aligned}\label{eq:dispersion}
\end{equation}
and $\hbar \omega_Z=\left(p_{0}+2\hbar k_{c}\right)^{2}/2M-p_0^2/2M$.

\subsection{Mean-field Approximation}
In this part, we derive  the mean-field equations of motion  for the field operator $\hat{\psi}^{\Uparrow (\Downarrow)}_{p_n}$, by approximating it as a c-number,  $\psi_{p_n}^{\Uparrow (\Downarrow)} \approx \left\langle \hat{\psi}^{\Uparrow (\Downarrow)}_{p_n} \right\rangle$. They are given by,
\begin{equation}
\begin{aligned}
        i\frac{d}{dt}\psi_{p_n}^{\Downarrow} &= \omega_{p_n}^{\Downarrow} \psi_{p_n}^{\Downarrow} + \chi N \psi_{p_n}^{\Uparrow} \sum_m (\psi_{p_m}^{\Uparrow})^* \psi_{p_m}^{\Downarrow}  \\
        i\frac{d}{dt}\psi_{p_n}^{\Uparrow} &= \omega_{p_n}^{\Uparrow} \psi_{p_n}^{\Uparrow}  + \chi N \psi_{p_n}^{\Downarrow} \sum_m (\psi_{p_m}^{\Downarrow})^* \psi_{p_m}^{\Uparrow} 
\end{aligned} \label{eq:mean-field}
\end{equation}
with $ |\psi_{p_n}^{\Downarrow}|^2 + |\psi_{p_n}^{\Uparrow}|^2 =1$ and $\psi_{p_n}^{\Downarrow} = \psi_{p_n}^{\Uparrow}=\frac{1}{\sqrt{2}}$ for the initial state, for the case of a $\pi/2$ pulse.

In the limit where $\chi N \gg 2k \sigma_p/m$,  atoms with different momentum lock together,  $\left\langle \hat{S}_X\right\rangle\approx N/2$ and  the exchange interaction simplifies to $\hat{H}_{\mathrm{ex}}= \chi N \hat{S}_X$. The exchange interactions acts effectively as  a strong transverse field along the initial spin direction. The corresponding  equations of motion  become:
\begin{equation}
\begin{aligned}
        i\frac{d}{dt}\psi_{p_n}^{\Downarrow} &= \omega_p^{\Downarrow} \psi_{p_n}^{\Downarrow} + \frac{\chi N}{2} \psi_{p_n}^{\Uparrow} \\
        i\frac{d}{dt}\psi_{p_n}^{\Uparrow} &= \omega_p^{\Uparrow} \psi_{p_n}^{\Uparrow} + \frac{\chi N}{2} \psi_{p_n}^{\Downarrow}.
\end{aligned} \label{eq:approx}
\end{equation}

After evolution under  $\hat{H}$ for time $t_d$, and replacing $p_n \to p$ for simplicity we get:
\begin{equation}
\begin{aligned}
    \sqrt{2}\psi^{\Downarrow}_{p}&=e^{-i\frac{p^{2}}{2M\hbar}t_{d}}\Bigr[\cos\left(\frac{t_{d}}{2}\sqrt{\left(2k p /M\right)^{2}+\left(\chi N\right)^{2}}\right)+\frac{i\left(2k p /M-\chi N\right)\sin\left(\frac{t_{d}}{2}\sqrt{\left(2k p /M\right)^{2}+\left(\chi N\right)^{2}}\right)}{\sqrt{\left(2k p /M\right)^{2}+\left(\chi N\right)^{2}}}\Bigr] \\
    \sqrt{2}\psi^{\Uparrow}_{p}&=e^{-i\frac{p^{2}}{2M\hbar}t_{d}}\Bigr[\cos\left(\frac{t_{d}}{2}\sqrt{\left(2k p /M\right)^{2}+\left(\chi N\right)^{2}}\right)-\frac{i\left(2k p /M+\chi N\right)\sin\left(\frac{t_{d}}{2}\sqrt{\left(2k p /M\right)^{2}+\left(\chi N\right)^{2}}\right)}{\sqrt{\left(2k p /M\right)^{2}+\left(\chi N\right)^{2}}}\Bigr].
\end{aligned}      
\end{equation}
Using the approximation
\begin{equation}
\frac{2k p /M\pm\chi N}{\sqrt{\left(2k p /M\right)^{2}+\left(\chi N\right)^{2}}}\approx\pm\mathrm{sgn}(\chi),
\end{equation} one obtains

\begin{equation}
    \sqrt{2}\psi^{\Downarrow}_{p} \approx \sqrt{2}\psi^{\Uparrow}_{p}  \approx
    e^{-i\frac{p^{2}}{2M\hbar}t_{d}}e^{-i\frac{t_{d}}{2}\sqrt{\left(\frac{2k_{c}p}{M}\right)^{2}+\left(\chi N\right)^{2}}}  \approx e^{-i\frac{\chi N}{2}t_{d}}e^{-i\left(\frac{1}{2M\hbar}+\frac{k_{c}^{2}}{\chi N M^{2}}\right)p^{2}t_{d}}.
\end{equation}
 At the   optimal interaction strength, $\chi_{\mathrm{opt}}$, 
\begin{equation}
\chi_{\mathrm{opt}} N = - 4 \frac{\hbar k^2}{2 M} = -4 E_R.
\end{equation}  the global phase is canceled,

\subsection{General initial polar angle}

In this section, we consider the case where the  initial state is prepared as a superposition between $\ket{p_0}$ and $\ket{p_0+2\hbar k}$ with an arbitrary  polar angle $\theta$, as described in the main text. In this case, the conserved total magnetization  $\left\langle \hat{S}_Z \right\rangle = \frac{N}{2}\cos\theta$. Under the mean-field approximation and  assuming dominant exchange interactions over  single-particle inhomogeneities $\left\langle \hat{S}_+ \right\rangle\approx \frac{N}{2}\sin\theta e^{-2i\chi \langle \hat{S}_Z\rangle t}$~\cite{norcia2018cavity}.

The mean-field Hamiltonian thus acquires a time-dependent form: 
\begin{equation} 
\hat{H}_{\rm MF}/\hbar = \hat{H}^d_{\rm in} /\hbar + \frac{\chi N \sin\theta}{2}\left(\hat{S}_{+}e^{i\chi N\cos\theta t} + \hat{S}_{-}e^{-i\chi N\cos\theta t}\right). 
\end{equation}
By transforming into a rotating frame defined by $\hat{H}_R=-\chi N\cos\theta\hat{S}_Z$, the Hamiltonian simplifies to: 
\begin{equation} 
\hat{H}'_{\rm MF}/\hbar= \sum_r\left(\frac{2k p_r}{M}+\chi N\cos\theta\right)\hat{s}_{p_r}^z + \chi N\sin\theta \hat{S}_X. 
\end{equation}
The relevant eigenstate in this rotating frame has the energy dispersion: 
\begin{equation} 
\begin{aligned} E_p &= -\frac{1}{2}\sqrt{\left(\chi N\cos\theta+\frac{2k p}{M}\right)^2+\left(\chi N\sin\theta\right)^2}+\frac{p^2}{2M\hbar}\\
&\approx \frac{\chi N}{2}-\frac{k p\cos\theta}{M}+\frac{p^2}{2M\hbar}\left(1+\frac{4E_R\sin^2\theta}{\chi N}\right).
\end{aligned} 
\end{equation}
The linear term in $p$ generates a shift in   collective recoil of the two wave packets, by $ -\frac{k \cos\theta}{M}$  and thus a net average momentum  of $p_0 + \hbar k (1-\cos\theta)$. Then the optimal interaction strength that cancels the dispersion is given by
\begin{equation} 
\chi_{\rm opt} N = -4E_R\sin^2\theta. 
\end{equation}

\section{Detection scheme for thermal solitons}
\begin{figure}
\includegraphics[width=1\columnwidth]{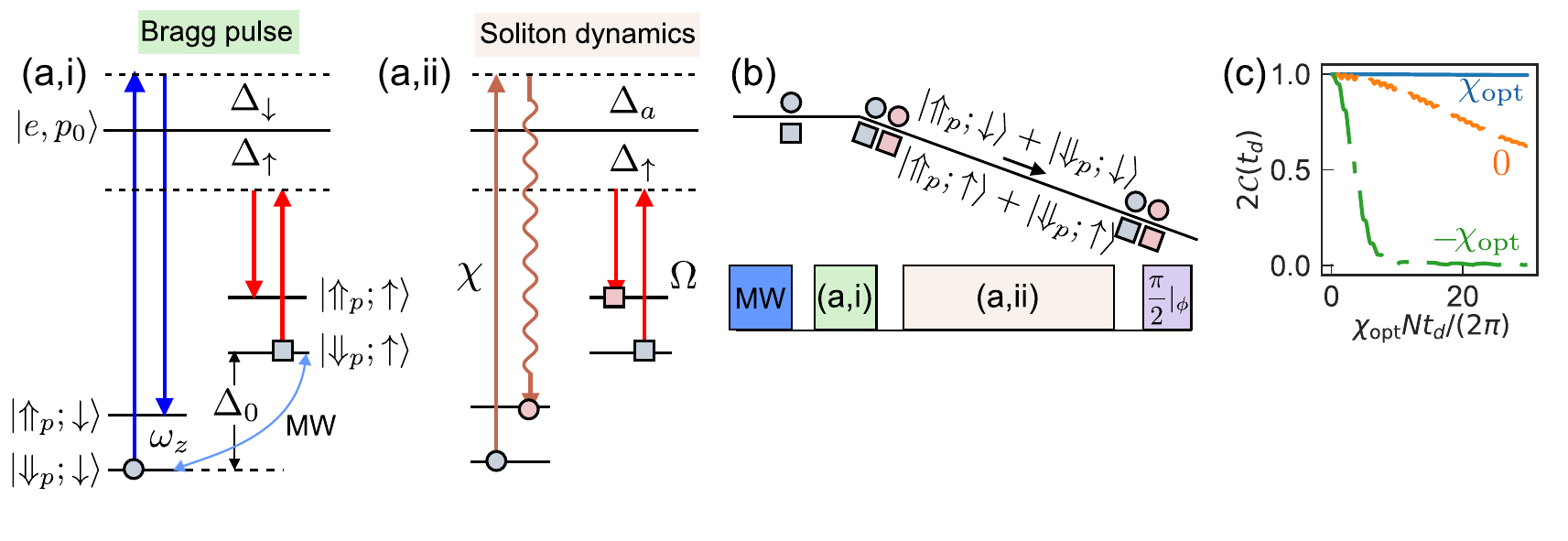}
\caption{\label{fig:detect} Detection scheme for a thermal cloud 
(a,i) A microwave drive couples two hyperfine levels with an energy difference $\Delta_0$ (light blue arrow). Two sets of Bragg beams (blue and red) independently prepare momentum superpositions in each of the hyperfine levels.(a,ii) Cavity photons  (orange) and  Bragg beams (red)  are used to control the motional dynamics for different hyperfine levels respectively.
(b) Interferometer sequence to measure wave-packet broadening.
(c) Simulated measurement outcome using a  Bragg pulse with Rabi frequencies,  $\Omega=\chi_{\rm opt}$ (red), $\chi=\chi_{\rm opt}$ (blue), $\chi=-\chi_{\rm opt}$ (orange) and $\chi=0$ (green).} 
\end{figure}
%%%%%%%%%%%%%%%%%%%%%%%%%%%%%%%%%%%%%%%%%%%%%%%%%%%%%%%%%%%%%%%%%%%%%%%%%%
We now discuss the detection scheme proposed in the main text for thermal solitons. In the following, we focus primarily on the 1D scenario. Soliton generation  in  2D and 3D systems already requires the introducion of  additional internal degrees of freedom to enhance the effective energy difference $\omega_Z$, and  therefore adding another state is possible but  more cumbersome. In the high-dimensional cases, therefore, it will be easier to start with a BEC and use standard time of flight images. 

The protocol relies on the fact that a similar flattening of the band can be realized for non-interacting atoms using  an additional driving field along the $x$-axis of the Bloch sphere with Rabi frequency $\Omega=-4E_R$.
As illustrated in Fig.~\ref{fig:detect}(a,ii), the key idea  then is to use an additional external drive (red arrows), instead of exchange interactions, to prepare a soliton between a pair of momentum states in an other  hyperfine level $\ket{\uparrow}$, and inject it into one arm of the interferometer. 
We choose the cavity frequency to be off-resonant from the atomic transition for this internal state, making it  insensitive to exchange interactions. In the other arm, we inject a soliton in a hyperfine state $\ket{\downarrow}$ that experiences exchange interactions (orange arrows). The contrast achieved at the output of the interferometer serves as a probe of the  differential dispersion. 

We adopt the notation $\ket{\Downarrow_p,\Uparrow_p;\downarrow,\uparrow}$ for the momentum (first index) and spin basis (second index), and ignore the atom number index $r$ for simplicity. 
The interferometer sequence, shown in the spacetime diagram of Fig.~\ref{fig:detect}(b), consists of the following steps:
\begin{itemize}
    \item After velocity selection $\ket{\Downarrow_p,\downarrow}$, a microwave pulse is used to create a superposition of internal states $\frac{1}{\sqrt{2}}(\ket{\Downarrow_p,\downarrow} + \ket{\Downarrow_p,\uparrow})$ (Fig.~\ref{fig:detect}(a,i) light blue arrow);
    \item Two sets of Bragg beams with opposite single-photon detunings $\Delta_{\downarrow(\uparrow)}$ from the excited state (Fig.~\ref{fig:detect}(a,i) dark blue and red arrows) generate a density grating within each internal state as $\frac{1}{2}(\ket{\Uparrow_p,\downarrow} + \ket{\Downarrow_p,\downarrow} + \ket{\Uparrow_p,\uparrow} + \ket{\Downarrow_p,\uparrow})$;
    \item  The blue-detuned Bragg beams are turned off and the red-detuned Bragg frequency is tuned to $\Omega=\chi_{\rm opt}$ (red) for the hyperfine level $\ket{\uparrow}$, while the cavity-mediated interactions generated by the probe beam (orange) are experienced by the hyperfine level $\ket{\downarrow}$ (Fig.~\ref{fig:detect}(a,ii));
    \item After some time evolution, a second set of  Raman  and a microwave  $\pi/2$  pulses is applied. The fringe  contrast reveals the overlap between the wave packets in the two arms.
\end{itemize}

In this way,  the spatial overlap between wave packets in the interferometer arms can be mapped onto the population difference at the interferometer output.

%The spatial overlap between wave packets in different interferometer arms onto the population difference of an ancillary internal state at the interferometer output, analogous to the celebrated Hadamard test in quantum information science. 

Figure .~\ref{fig:detect}(c)) shows an example of the signal  performed by varying the  exchange interaction strength $\chi N$ for $\ket{\downarrow}$ while setting  the Raman drive Rabi frequency to $\Omega=\chi_{\rm opt}$: Three cases are used,  $\chi_{\rm opt}$ (blue), $0$ (green), and $-\chi_{\rm opt}$ (orange). The simulations reveal distinct behaviors in each case: for $\chi=\chi_{\rm opt}$, both of the  $\ket{\uparrow}$ and $\ket{\downarrow}$ momentum superpositions exhibit perfect soliton dynamics, maintaining constant overlap; with $\chi=0$, the wave packets for $\ket{\downarrow}$ spatially separate due to lack of gap protection;  in this case an echo sequence should help to increase the overlap but will not be perfect due to broadening of one of the superpositions. For $\chi=-\chi_{\rm opt}$, the wave packets for $\ket{\downarrow}$ remain bound but broaden, and the contrast can decay faster than the non-interacting case.

\section{Balanced pump scheme}
In the prior discussions, we focused on ideal unitary dynamics that excluded dissipation. However, Ref. \cite{luo2023cavity,luo2025hamiltonian} demonstrated that superradiance is a detrimental effect, that can drive the  state toward either the north or south pole, disrupting  coherence and reducing  the gap protection. 

\begin{figure}
\includegraphics[width=0.5\columnwidth]{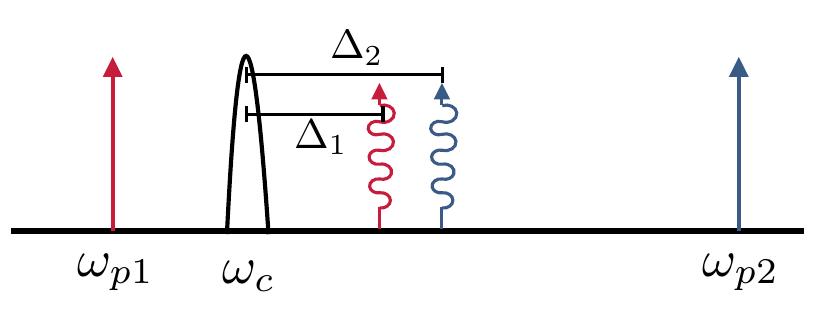}
\caption{\label{fig:squeeze} A two pump scheme can be used  to cancel superradiance and maintain a dominant unitary dynamics.}
\end{figure}

A straightforward strategy to mitigate superradiance is to detune the pump field further from the cavity resonance. However, this approach results in slower dynamics that can be then  significantly affected by light scattering. An alternative method is illustrated in Fig.~\ref{fig:squeeze}, where the cavity is driven by two distinct frequency tones. Unlike the dual-pump resonance condition in \cite{luo2025hamiltonian,luo2024realization}, here we choose $\Delta_1 \neq \Delta_2$. In this setup, the two pumps independently generate exchange interactions, described by:
\begin{equation}
\begin{aligned}
\hat{H} &= \chi_1 \hat{S}_{-} \hat{S}_{+} + \chi_2 \hat{S}_{+} \hat{S}_{-} \approx (\chi_1+\chi_2) \left(\hat{S}^2 - (\hat{S}_Z)^2 \right),\\
\hat{L}_1 &= \sqrt{\Gamma_1} \hat{S}_{+}, \quad \hat{L}_2 = \sqrt{\Gamma_2} \hat{S}_{-},
\end{aligned}
\end{equation}
and 
\begin{equation}
\Gamma_1 = \left(\frac{g^2}{4\Delta_0}\right)^2 |\alpha_1|^2 \frac{\kappa}{\Delta_1^2 + (\kappa/2)^2} \quad \Gamma_2 = \left(\frac{g^2}{4\Delta_0}\right)^2 |\alpha_2|^2 \frac{\kappa}{\Delta_2^2 + (\kappa/2)^2}.
\end{equation}
By setting $\Gamma=\Gamma_1=\Gamma_2$,
it is possible  to suppress superradiance at the mean-field level. To be more concrete, the mean-field equation of motion get  modified in the balanca pump case to 
\begin{equation}
\begin{aligned}
\frac{d}{dt} \left\langle \hat{S}_X \right\rangle&=\cdots - \Gamma \left\langle \hat{S}_X \right\rangle\\
\frac{d}{dt}  \left\langle \hat{S}_Y \right\rangle&=\cdots - \Gamma  \left\langle \hat{S}_Y \right\rangle\\
\frac{d}{dt}  \left\langle \hat{S}_Z \right\rangle&=\cdots - 2\Gamma  \left\langle \hat{S}_Z \right\rangle\\
\end{aligned}
\end{equation}
with $\cdots$ representing the contributions from the unitary dynamics. 
Note that due to the balance in the superradiance processes, the collective enhancement for superradiance disappears, resulting in a much slower time scale (by $O(N)$) compared to the collective dynamics.

\bibliographystyle{unsrt}
\bibliography{reference}